\documentclass[aip,reprint]{revtex4-1}
\usepackage{graphicx}
\usepackage{dcolumn}
\usepackage{bm}

\begin{document}

\title{Electronic Thermal Conductivity Measurements in Intrinsic Graphene} 

\author{S. Yi\u{g}en, V. Tayari, J.~O. Island, J. M. Porter, A.~R. Champagne}
\email[]{a.champagne@concordia.ca}
\affiliation{Department of Physics, Concordia University, Montr\'{e}al, Qu\'{e}bec, H4B 1R6 Canada}

\date{\today}

\begin{abstract}
The electronic thermal conductivity of graphene and 2D Dirac materials is of fundamental interest and can play an important role in the performance of nano-scale devices. We report the electronic thermal conductivity, $K_{e}$, in suspended graphene in the nearly intrinsic regime over a temperature range of 20 to 300 K. We present a method to extract $K_{e}$ using two-point DC electron transport at low bias voltages, where the electron and lattice temperatures are decoupled. We find $K_e$ ranging from 0.5 to 11 W/m.K over the studied temperature range. The data are consistent with a model in which heat is carried by quasiparticles with the same mean free-path and velocity as graphene's charge carriers.
\end{abstract}

\pacs{} \keywords{}

\maketitle 

The electronic heat conductivity of graphene, $K_{e}$, describes how charged quasiparticles carry energy as they diffuse in this material. It could also shed light on $K_{e}$ in other 2D Dirac systems whose electronic band structure is related to graphene's, such as the surface states of topological insulators\cite{Butler13}. When a hot electron diffuses out of graphene, it cools down the electronic distribution. Thus, measurements of $K_{e}$ are needed to complement the understanding of the other hot-electron cooling mechanisms in graphene which involve various electron-phonon couplings\cite{Bistritzer09, Tse09, Kubakaddi09, Berciaud10, Efetov10, Song12, Betz12, Betz13, Graham13,DasSarma11}. Measuring and controlling $K_{e}$ could have applications in the heat management of heavily-doped nm-scale devices where $K_{e}$ can be dominant\cite{Saito07}, and in optimizing graphene's electro-optical properties\cite{Gabor11,Song11}. While there have been several experimental reports of the phononic thermal conductivity, $K_{p}$, in graphene \cite{Ghosh08,Freitag09,Seol10, Gabor11, Jo11, Balandin11, Pop12, Dorgan13}, there is no report of $K_{e}$ measurements in suspended graphene. This is because in most regimes $K_{p}$ is much larger than $K_{e}$, which makes it difficult to measure the amount of heat carried by the charged quasiparticles (electron and holes).

We present a carefully calibrated method to extract $K_{e}$ in graphene using DC electron transport in suspended devices. The accuracy of the method is dependent on high-mobility (annealed) devices. We present data from three different samples which show consistent results. The extracted $K_e$ are compared with calculated values, $K_{e-th}$, for a diffusing gas of Dirac quasiparticles. The agreement between theory and measurements is quantitative for all three devices over the temperature range (20 - 300 K) studied. Throughout the text we use $T$ to designate the lattice (cryostat) temperature, and $T_{e}$ for the average electron temperature in the suspended devices. At very low bias, $|V_{B}| \lesssim$ 1 mV, $T = T_{e}$. We first describe our samples, secondly we present our $T_{e}$ thermometry, then show how we apply a controlled $\Delta T$ using Joule heating, and finally extract $K_{e}$ from the transport data.

\begin{figure}
\includegraphics [width=3.25in]{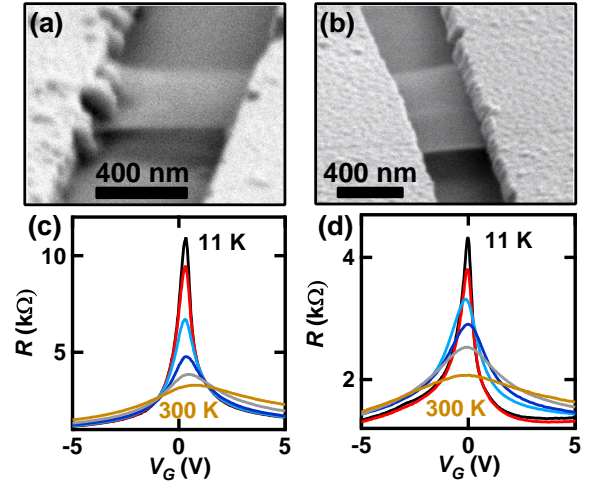}
\caption{\label{}(Color online.) Suspended graphene devices. (a) and (b) tilted SEM images of 650 and 400 nm long suspended graphene transistors (Samples A and B). (c) and (d) $R$ of Samples A and B versus gate voltage, $V_G$, at $T_{e} = T$ = 11, 50, 100, 150, 210 and 300 K, and $V_{B} =$ 0.5 mV.}
\end{figure}
Figure 1 (a)-(b) shows tilted SEM images of Sample A and B respectively (Sample C in Supplemental Material\cite{SM}, SM, Fig.\ S1). We confirmed using optical contrast and Raman spectroscopy that all three samples are single-layer graphene. Sample A is 650 nm long, 675 nm wide, and suspended 140 $\pm$ 10 nm above the substrate (AFM measurement) which consists of 100 $\pm$ 2 nm of SiO$_2$ (ellipsometry measurement) on degenerately-doped Si which is used as a back-gate electrode. Sample B is 400 nm long, 1.05 $\mu$m wide, and suspended 175 $\pm$ 10 nm above a 74 $\pm$ 2 nm SiO$_2$ film on Si. To prepare the samples, we used exfoliated graphene, and standard e-beam lithography to define Ti(5nm)/Au(80nm) contacts. The samples were suspended with a wet BOE etch such that their only thermal connection is to the gold contacts. We annealed the devices using Joule heating \textit{in situ} by flowing a large current in the devices\cite{Bolotin08} (up to 540, 840 and 837 $\mu$A for A, B and C). Annealing and subsequent measurements were done under high vacuum,$~ 10^{-6}$ Torr.

Panels (c) and (d) show DC two-point resistance data, $R = V_{B}/I$, for Samples A and B respectively after annealing, versus gate voltage, $V_{G}$, which controls charge density. From the width of the $R$ maximum at 11 K, we extract a half-width-half-maximum, HWHM, of 0.45, 0.6 and 0.95 V for Samples A, B, C (Sample C, Fig.\ S1). Using a parallel plate model for the gate capacitance of the devices, these HWHMs correspond to an impurity induced charge density\cite{Du08} of $n^{*} \approx$ 1.5, 1.7 and 2.1 $\times 10^{10} $ cm$^{-2}$.
\begin{figure}
\includegraphics[width=3.25in]{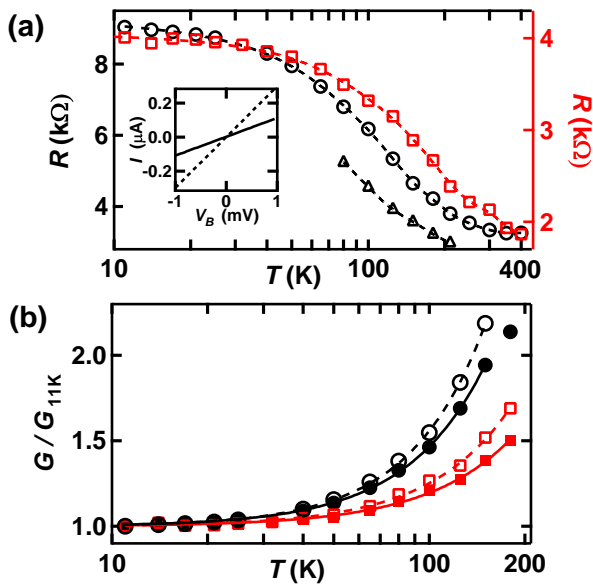}
\caption{\label{}(Color online.) Electron thermometry. (a) Temperature dependence of $R$ in Sample A (circles, left axis), Sample B (squares, right axis), and Sample C (triangles, left axis) near the charge degeneracy $n_{G} =$ 5.7, 2.9 and -1.5 $\times 10^{9}$ cm$^{-2}$. The dashed lines are numerically interpolated curves used for thermometry. Inset: $I-V_{B}$ data for Sample A, $|V_{B}| < 1$ mV, whose slope is used to extract $R$. The solid (dashed) line is at $T = T_{e}$ = 11 (300) K. (b) Relative conductance $G/G_{11 K}$ of Sample A (circles) and Sample B (squares) versus $T  = T_{e}$. The filled symbols show the raw two-point data, and the open symbols the data after subtracting $R_c = R_{o-Dirac}$ (see text). The solid and dashed lines are power law fits consistent with charge impurity scattering.}
\end{figure}

The devices were fabricated with large contact areas between the gold electrodes and graphene crystals, 1.1 to 3 $\mu m^2$ per contact, to minimize the contact resistance, $R_{c}$. An upper bound for $R_{c}$ can be extracted from the two-point $R-V_{G}$ curve in Fig. 1(c) by fitting the data (SM\cite{SM} section 2) with the expression\cite{Castro10} $R = R_o + (L/W)(1/n_{G}e\mu)$ where $R_o$ is the resistance due to neutral scatterers plus $R_{c}$, $L$ is the length of the device, $W$ the width, $n_{G}$ the charge density induced by $V_{G}$, $\mu$ the mobility, and $e$ the electron's charge. We fit the data at $T = T_{e} =$ 100 K for ($V_{G}-V_{D}$) $>$ 1.3 V to avoid the thermal smearing around the Dirac point, $V_D$. The extracted mobility for Sample A in the doped regime is $\mu \approx 8.5 \times 10^{4}$ cm$^{2}$/V.s at 100 K, and $R_{o} \approx$ 682 $\pm$ 53 and 1135 $\pm$ 80 $\Omega$ for hole and electron doping respectively. The difference between hole, $R_{o-h}$ and electron doping, $R_{o-e}$, is understood as an additional $p-n$ barrier for the electron due to $p$-doping from the gold electrodes \cite{Castro10}. At the Dirac point, we let $R_{o-Dirac} = (R_{o-h}+R_{o-e})/2 =$ 908.5 $\Omega$ for Sample A. For Sample C, we find $R_{o-Dirac} =$ 1097 $\Omega$. We note that $R_{o-Dirac}$ is much smaller than $R$ of Samples A and C, therefore $R_{c} < R_{o-Dirac}$ has at most a modest impact on our measurements in these devices. It is not possible to extract $R_o$ for Sample B because it enters the ballistic regime away from the Dirac point (doped regime)\cite{SampleB}.  The contact areas of Sample B are larger, and its width wider, than for Samples A and C. Assuming a similar resistance per unit area as for A and C, we expect $R_{c} \lesssim$ 657 $\Omega$ for B. Based on the reported thermal conductance of Au/Ti/Graphene and Graphene/SiO$2$ interfaces\cite{Koh10}, the thermal resistance of our contacts are several orders of magnitude lower than the one we measure below for graphene. Thus, the thermal resistance of the contacts can safely be neglected.

Figure 2(a) shows $R$ vs cryostat temperature, $T$, calibration curves for Samples A (circles, left axis), B (squares, right axis), and C (triangles, left axis) near $V_{G} = V_{D}$. $R = V_{B}/I$ data are extracted from the slope of the $I-V_{B}$ data as shown in the inset of panel (a) at 11 K (solid) and 300 K (dashed), for $\pm$ 1 mV bias where no Joule heating effect is present ($T_{e} = T$). The data are taken at $V_{G} =$ 0.5 V close to $V_{D} =$ 0.33 V for Sample A, and at $V_{G}=$ 0 V for Samples B and C ($V_{D} =$ -0.1 and 0.07 V), corresponding to $n_{G} =$ 5.7, 2.9, and -1.5 $\times 10^{9}$ cm$^{-2}$. The $T$ dependence of the data shows an insulating behavior up to $\approx $ 200 K for Sample A and C, and up to 300 K for Sample B. The interpolated dashed lines in panel (a) will be used as thermometry curves to monitor $T_e$. Note that the thermometry is most accurate where the curves are steepest.

Figure 2(b) shows the relative conductance $G(T)/G_{11 K}$ in the intrinsic regime extracted from panel (a) for Sample A and B. The $T$ dependence of $G$ in graphene, at low charge density, is strongly dependent on the type of charge transport. For ballistic transport, we expect a very weak temperature dependence at low $T$, and a linear dependence when $k_{B}T >> E_{F}$ \cite{Muller09}. In the diffusive regime, the expected temperature dependence depends on the type of charge scatterers, and $G(T)/G_{11 K} \propto T^{\alpha}$ with $\alpha = $ -1, 0, 2 for acoustic phonon, short-range (neutral), and long-range (charged) scatterers respectively \cite{DasSarma12, DasSarma11}. The temperature dependence of real samples is expected to combine all three types of scattering. We fit the data with a function $G/G_{11 K} = 1 + AT^{p}$, and extract $p$ = 1.85, 1.74, 1.72 and 1.63 $\pm$ 0.03 for Sample A with $R_{c} = R_{o-Dirac}$ and 0 (open and filled circles), and Sample B with $R_{c} =$ 657 and 0 $\Omega$ (open and filled squares). This $T$-dependence strongly supports diffusive charge transport dominated by long-range charge impurities, as reported in previous experiments on high-mobility devices \cite{Bolotin08,DasSarma11} and expected theoretically \cite{DasSarma12}. The small departure from a $T^{2}$ dependence is expected as the samples are not exactly at the Dirac point. We conclude that all samples are in the diffusive regime at low charge density (Fig. 2(b) and SM section 3) and scattering is predominantly due to charged impurities. The data in Fig. 2(a), and its agreement with theory, serves as a reliable thermometer for $T_{e}$ in our devices.

\begin{figure}
\includegraphics[width=3.25in]{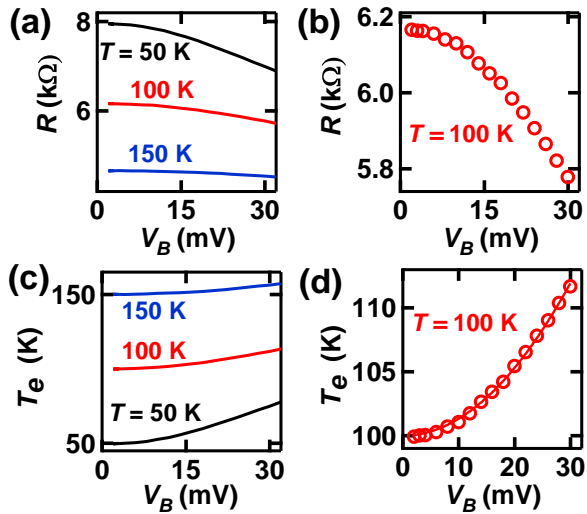}
\caption{\label{}(Color online.) Electron heating. (a) $R$ vs $V_B$ at various $T$ for Device A at $V_{G} =$ 0.5 $V \approx V_{D}$. (b) Details of the data at $T =$ 100 K. Joule heating due to $V_B$ raises the flake's average $T_e$ above $T$. $T_{e}$ is extracted using Fig. 2(a). Panels (c) and (d) show $T_e$ vs $V_{B}$ in Sample A at a few different $T$, and at $T =$ 100 K respectively.}
\end{figure}

After establishing the $T_{e}$ thermometry, we demonstrate controlled Joule self-heating of the electrons to apply a thermal bias $\Delta T = T_{e} - T$ between the suspended graphene and the electrodes. Figure 3(a) shows $R$ vs $V_{B}$ for Sample A at $T = $ 50, 100, 150 K (for Samples B and C see Figs.\ S3 and S4). Panel (b) shows the details of the data at 100 K. $R$ decreases monotonically with increasing $V_{B}$, at all $T$. We argue that this change in the $R$ vs $V_{B}$ data is caused by Joule heating of the sample. Other mechanisms which could cause a non-linear $I-V_{B}$ relation include: scattering from flexural phonons, in-plane optical phonons, substrate phonons, and Zener-Klein tunneling. We restrict our measurements to $V_{B}$ $\lesssim 30$ meV. This rules out any $R$ change due to scattering from optical in-plane phonons, $\approx$ 200 meV, and flexural phonons, $\approx$ 70 meV, in graphene\cite{DasSarma11}. Phonons in the substrate can also be ruled out as the samples are suspended. The contribution of Zener-Klein tunneling to $I - V_{B}$  non-linearity was only observed in very low-mobility devices, and at $V_{B} > 100$ mV \cite{Vandecasteele10}. This leaves Joule heating as the only plausible cause for the observed $R$ vs $V_{B}$ behavior \cite{Viljas11}. Using the calibration curve for the samples, Fig.\ 2(a), and data from Fig. 3(a)-(b), we extract the average $T_{e}$ vs $V_{B}$, as shown for Sample A in Fig.\ 3(c)-(d). In Fig. 3(d), we fit a power law (solid line) $T_e = 100 + BV_{B}^{x}$, and find $x=$ 2.00 $\pm$ 0.04, as expected for Joule heating over a small $T_e$ range where $K_{e}$ and $R$ do not change appreciably (Samples B and C, SM section 4). Figs.\ 3(d), S3(d), S4(d) show that the accuracy with which $T_{e}$ can be extracted is much better than 1 K. We calculate $T_{e}$ errors from the scatter of the data in panel (d), and similar plots at each $T$, to vary from 0.2 K (steepest regions of Fig.\ 2(a)) up to 2 K (flat regions of Fig.\ 2(a)). The smooth dependence of $T_{e}$ on $V_{B}$ at all $T$ is consistent with electrons having a well defined temperature, as predicted by calculations of the $e-e$ collision length\cite{Li13} (SM section 5). This is also confirmed by the $K_{e}$ data shown below.

Since our devices are much wider than the elastic mean-free path (SM section 3), the effect of their edges on transport should be small. We use a 1-d heat equation to extract $K_{e}$ in our devices: $K_{e}\frac{d^{2}T_{e}}{dx^{2}}+ Q = 0$ where $Q = RI^{2}/WLh$ is the Joule heating power per unit volume, $W$ the width, $L$ the length, and $h$ = 0.335 nm the thickness. Using boundary conditions $T_{e}= T$ at the two ends (contacts) of the flake, we find $T_{e}(x) = T + (LQx - Qx^{2})/2K_{e}$. Averaging over the length we find, $T_{e} = (1/L)\int_0^{L} T_{e}(x)$ $dx$ $= T + (QL^2)/(12K_{e})$. Finally, $K_{e} = \frac{QL^{2}}{12\Delta T}$, where $ \Delta T = T_{e} - T$. Using $R$ and $I$ from Fig. 3 and similar plots, for $\Delta T$ = 1, 2 and 5 K we extract $K_{e}$ vs $T_{e}$ in Fig. 4(a) for Sample A. Panel (b) shows $K_{e}$ vs $T_{e}$ for all three samples for $\Delta T$ = 5 K. Data in Fig. 4 show a strong $K_{e}$ dependence on $T_{e}$ ranging from roughly 0.5 W/K.m at 20 K to 11 W/K.m at 300 K. The $T_e$ range is limited to the region where we have accurate thermometry (Fig. 2(a)), up to $\approx$ 200 K for A and C, and 300 K for B. Error bars representing the total uncertainty on $K_{e}$ are shown for the $\Delta T =$ 5 K data (see SM section 6). If the $V_{B}$ needed to apply $\Delta T$ were to dope significantly the samples, it could affect the measured $K_{e}$. Using $n_{tot}(T)$ (SM section 3)\cite{Dorgan10}, we define an effective chemical potential $\mu_{eff}(T) = \hbar v_{F}\sqrt{\pi n_{tot}(T)}$. For instance, at $T$ = 100 K, $\mu_{eff}(100K) =$ 18, 18.4 and 19.5 meV respectively for the three devices. The $V_{B}$ necessary to achieve $\Delta T \leq$ 5 K in Fig. 4 is always smaller than $\mu_{eff}(T)$. We only observe a change in the extracted $K_{e}$ values when $\Delta T$ exceeds 8 K, and $V_{B} > \mu_{eff}(T)$. Thus $V_{B}$ does not affect our $K_{e}$, with the caveat that we cannot extract $K_{e}$ precisely at $n=$ 0. The thermoelectric voltages in our devices are negligible compared to $V_{B}$\cite{Zuev09, Hwang09}.
\begin{figure}
\includegraphics[width=3.25in]{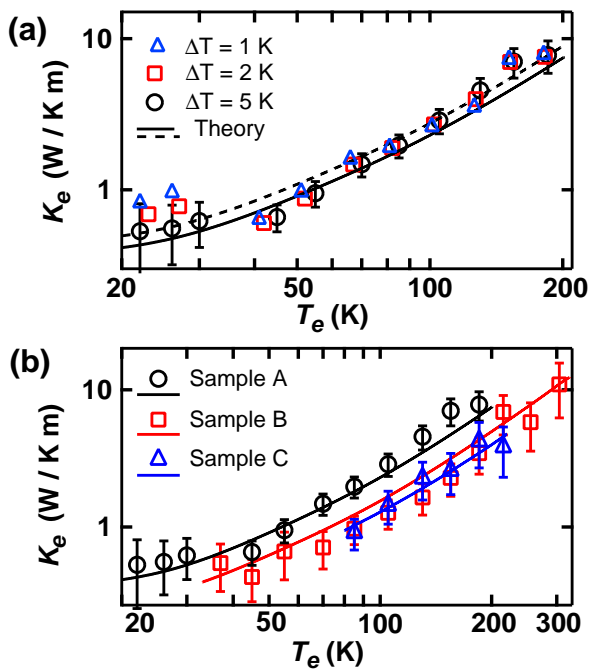}
\caption{\label{}(Color online.) Electronic thermal conductivity, $K_{e}$, in the quasi-intrinsic regime, $n_{tot, T=0}\approx$ 1.7 (Samples A, B) and 2.1 (Sample C) $\times 10^{10}$ cm$^{-2}$. (a) $K_{e}$ vs $T_{e}$ for $\Delta T = T_{e} - T =$ 1, 2  and 5 K for Sample A. The solid line is a theoretical calculation of $K_{e-th}$. The dashed line shows the same calculation with a contact resistance $R_c = R_{o-Dirac}$. The error bars are shown for the $\Delta T =$ 5 K data. (b) $K_{e}$ vs $T_{e}$ for $\Delta T =$ 5 K for Samples A, B and C, and $K_{e-th}$ for each sample.}
\end{figure}

We compare our data with the usual model for diffusing particles in 2-dimensions, $K_{e-th} = \frac{1}{2}Cvl$. If the heat flow is due to charge carriers, then the specific heat is $C = C_{e}$, the velocity is $v_{F} = 10^{6}$ m/s, and the mean free path $l$ is the same as for charge transport. We find (SM section 3), $l_{A-avg}$, $l_{B-avg}$, and $l_{C-avg}$ = 71 (85), 47 (59), and 37 (51) nm on average over the $T_{e}$ range with $R_{c}=$ 0 ($R_{o-Dirac}$). We calculate $C_{e}$ using the density of states for graphene and the Fermi-Dirac distribution (SM section 7). We plot $K_{e-th}$ in Fig. 4(a)-(b) as solid lines with $R_{c} =$ 0. They capture the quantitative $T_{e}$ dependence of our $K_{e}$ data. The $K_{e}$ data points are in good quantitative agreement with the calculated values for all three samples, and especially for Samples A and C. The dashed line in panel (a) shows $K_{e-th}$ if we use $R_c = R_{o-Dirac}$. If we account for $R_{c}$, i.e smaller $Q$, $K_{e}$ changes by the same magnitude as $K_{e-th}$ but in the opposite direction (not shown for clarity). The quantitative agreement between data and theory is not as accurate for Sample B since its $R$ is smaller than for Samples A and C, and the impact of $R_{c}$ could be bigger. The data and calculations shown in panel (b) with $R_{c}=$ 0 are within 20, 30, and 15 $\%$ of each other for Samples A, B and C. If we include $R_{c} = R_{o}$, which overestimates the effect due to $R_{c}$, the agreement between the data and theory for Sample B is at worst within a factor of two, and much better for Samples A and C. We fit a power law expression $K_{e} \propto T^{p}$ over $T_{e}$ = 45 - 185 K for Samples A and B, and find $p$ = 1.73 $\pm$ 0.15 and 1.63 $\pm$ 0.13 which is very close to the fit on $K_{e-th}$, $p_{th}$ = 1.62 and 1.59. This agreement is preserved even if we let $R_{c} = R_{o}$. As expected $p_{th}$ goes to 2 when $\mu/kT \ll 1$. We conclude that the $K_{e}$ data is consistent with heat being carried by particles moving with the $v_{F}$ and $l$ of the charge carriers. The magnitude of $K_{e}$ reaches $\approx$ 11 W / K.m at 300 K with $n_{tot, T=0}\approx$ 1.7 - 2.1 $\times 10^{10}$ cm$^{-2}$.

A condition to make reliable $K_{e}$ measurements is that all of the Joule heat remains in the carriers until they diffuse to the leads. Both experiments and theory confirm that the electron-phonon energy transfer in high mobility graphene, at low $V_{B}$, is very small below 300 K \cite{Gabor11,Song11,Viljas11,DasSarma12}, and decreases at lower $T$ and $n$. In our devices, we extract a cooling length for hot electrons (SM section 8), $\xi \approx$  100 to 10 $\mu$m for $T_{e} =$ 20 to 300 K. Since $\xi$ is much longer than $L$, and $V_{B}$ below the energy of optical phonons, we expect $T_{e}$ and $T$ to be decoupled in our devices when $V_{B} \neq$ 0, and all of the Joule heat to be carried to the contacts by charge carriers. Indeed, the $K_{e}$ we measure are two to three orders of magnitude lower than the reported phonon thermal conductivity $K_{p}$ in graphene \cite{Balandin11,Pop12}.

In summary, we fabricated high quality suspended graphene devices, developed self-thermometry and self-heating methods to extract and control $T_{e}$, and the electronic thermal conductivity in graphene. We extracted $K_e$ in the quasi-intrinsic regime, $n_{tot, T=0}\approx$ 1.7 - 2.1 $\times 10^{10}$ cm$^{-2}$, from $T_e =$ 20 K to 300 K. The $K_e$ data in three different devices are in very good agreement with a model where heat is carried by diffusing Dirac quasiparticles. Our results provide evidence that the dominant electron cooling mechanism in intrinsic sub-micron graphene devices below 300 K is hot-electron diffusion to the leads. The theoretical model we use naturally leads to the Wiedemann-Franz relation in the doped-regime and suggests that it should be obeyed in graphene. We thank Andrew McRae for discussions. This work was supported by NSERC, CFI, FQRNT, and Concordia University. We made use of the QNI cleanrooms.


\begin{thebibliography}{35}
\bibitem [{Sam()}]{SampleB}%
  \BibitemOpen
  \href@noop {} {\bibinfo  {journal} {In the doped regime (away from the Dirac
  point) and at low temperature, the R vs T data for Sample B are consistent
  with ballistic electron transport. These data will be discussed in a future
  publication}\ }\BibitemShut {NoStop}%
\end{thebibliography}
\end{document}